\documentclass[twocolumn,superscriptaddress,notitlepage,nofootinbib]{revtex4-2}
\usepackage{graphicx}
\usepackage{amsmath,amssymb,setspace}
\usepackage{newtxtext}
\usepackage{newtxmath}
\usepackage[x11names]{xcolor}
\usepackage[unicode,colorlinks,citecolor=Blue3,linkcolor=Blue3,bookmarks=true]{hyperref}

\newcommand{\eg}{{\it e.g.\,}}

\renewcommand{\Im}{\mathop{\rm Im}\nolimits}
\newcommand{\fulld}[2]{\dfrac{d#1}{d#2}}

\begin{document}

\title{Intracavity squeezing for Kerr QND Measurement scheme}

\date{Draft of \today}

\author{Dariya Salykina}%
\email{koi257@mail.ru}
\affiliation{Faculty of Physics, M.V. Lomonosov Moscow State University, Leninskie Gory 1, Moscow 119991, Russia}
\affiliation{Russian Quantum Center, Skolkovo IC, Bolshoy Bulvar 30, bld.\ 1, Moscow, 121205, Russia}

\author{Stepan Balybin}
\affiliation{Faculty of Physics, M.V. Lomonosov Moscow State University, Leninskie Gory 1, Moscow 119991, Russia}
\affiliation{Russian Quantum Center, Skolkovo IC, Bolshoy Bulvar 30, bld.\ 1, Moscow, 121205, Russia}

\author{Farid Ya.\ Khalili}
\email{farit.khalili@gmail.com}
\affiliation{Russian Quantum Center, Skolkovo IC, Bolshoy Bulvar 30, bld.\ 1, Moscow, 121205, Russia}

\begin{abstract}
  In Ref.\,\cite{22a1BaSaKh}, the scheme of quantum non-demolition measurement of optical quanta that uses a resonantly enhanced Kerr nonlinearity in the optical microresonator,  pre-squeezing of the probe beam, and its parametric amplification before the detection, was analyzed theoretically. It was shown that the main factor that limits the sensitivity of the considered scheme is the interplay of optical losses and the non-linear self-phase modulation (SPM) effect.

  Here we show that using the intracavity squeezing of the probe beam in this scheme, it is possible to cancel out the SPM effect. In this case, the sensitivity will be limited only by the available power in the pump beam and by the input/output losses in the signal beam. Our estimates show, that using the best optical microresonators currently available, the single-photon sensitivity for the intracavity photon number can be achieved. Therefore, this scheme could be of interest for optical quantum information processing tasks.
\end{abstract}

\maketitle

\section{Introduction}

The concept of quantum non-demolition (QND) measurements was formulated in late 1970s \cite{77a1eBrKhVo, Thorne1978, Unruh_PRD_19_2888_1979, 80a1BrThVo}. This type of measurement allows, in principle, to measure one given observable without perturbing it, thus implementing von Neumann's quantum reduction postulate \cite{Neumann_e}. The sufficient condition of such a measurement is commutativity of the measured observable with the Hamiltonian of the interaction of the explored object and the meter \cite{92BookBrKh, 96a1BrKh}.

In the case of the photon number measurement, the required Hamiltonian can be implemented using two optical beams, the signal and the probe ones, interacting through the Kerr ($\chi^{(3)}$) optical nonlinearity, see Refs.\cite{81a1eBrVy, Milburn_PRA_28_2065_1983, Roch_APB_55_291_1992, 96a1BrKh, Grangier_Nature_396_537_1998}. As a result, the phase of the probe beam is changed proportionally to the power of the signal beam and vice versa due to the cross-phase modulation effect (XPM) originating from the optical nonlinearity. The subsequent interferometric measurement of the probe beam phase gives information on the signal beam optical power, while perturbation of the signal beam phase by the probe beam power fulfills the Heisenberg uncertainty relation for this measurement. In the ideal lossless case, the photon numbers in both beams are preserved.


Unfortunately, the Kerr nonlinearity in highly transparent optical media is weak. This problem can be alleviated by using the high-$Q$ optical microresonators \cite{Strekalov_JOptics_18_123002_2016} that combine very low optical losses with a high concentration of the optical energy in the small volumes of their optical modes. The corresponding measurement scheme was considered in Ref.\,\cite{21a1BaMaKhStIlSaLeBi}, taking into account the optical losses. It was shown that the main factor that limits the sensitivity of this scheme is the interplay of the optical losses and the perturbation of the probe mode phase by its own power fluctuations (the self-phase modulation effect, SPM), and that the corresponding sensitivity limit is equal to
\begin{equation}\label{dNs_min_0}
  (\Delta N_s)^2 = \frac{\Gamma_S}{\Gamma_X^2}\sqrt{\frac{1-\eta}{\eta}} \,.
\end{equation}
Here $N_s$ is the photon number in the signal mode, $\Delta N_s$ is the measurement error, $\Gamma_X$, $\Gamma_S$ are the dimensionless factors of, respectively, the cross- and the self-phase modulation, see Eq.\,(6) and (50) of \cite{21a1BaMaKhStIlSaLeBi}, and $\eta$ is the quantum efficiency of the scheme (equal to one in the ideal lossless case). 

It was assumed in Ref.\,\cite{21a1BaMaKhStIlSaLeBi} that the initial state of the probe mode is the coherent one. However, it is known that the sensitivity of optical interferometric measurements can be improved by using non-classical squeezed states of light with the decreased uncertainty in the phase quadrature and anti-squeezing (parametric amplification) of the phase quadrature of the output beam before detection, as it was first proposed by C.\,Caves in Ref.\,\cite{Caves1981}. Currently, squeezed light is used in laser gravitational wave detectors \cite{Tse_PRL_123_231107_2019_short, Acernese_PRL_123_231108_2019_short}; the sensitivity gain provided by the output anti-squeezer was demonstrated experimentally in Ref.\, \cite{20a1FrAgKhCh} (see also the review \cite{22a1SaKh}).

In Ref.\,\cite{22a1BaSaKh}, the use of a squeezed state of light in the scheme of  Ref\,\cite{21a1BaMaKhStIlSaLeBi} was analyzed theoretically. It was shown that in this case, Eq.\,\eqref{dNs_min_0} takes the following form:
\begin{equation}\label{dNs_min_1}
  (\Delta N_s)^2 = \frac{\Gamma_S}{\Gamma_X^2}\sqrt{\frac{1-\eta}{\eta}}e^{-r-R} \,,
\end{equation}
where $r$ and $R$ are the logarithmic factors of the input squeezing and output anti-squeezing, respectively. It was shown in that work that the squeezing could allow to reach the QND measurement sensitivity of about a few tens of photons.

The limits \eqref{dNs_min_0} and \eqref{dNs_min_1} do not depend on the optical energy in the probe mode. Actually, they correspond to an optimal value of this energy. Therefore, the use of the most straightforward sensitivity resource, namely sending more optical power into the probe mode, is limited in both these cases.

Yet another squeezing technique is the intracavity squeezing, that is squeezing the light directly inside the optical cavity. It was proposed in Ref.\,\cite{Peano_PRL_115_243603_2015} as a method of suppressing of the quantum noise in optomechanical position sensors. A more detailed analysis of this technique, taking into account the optical losses, can be found in Ref.\,\cite{Korobko_PRA_108_063705_063705}. Recently, this method was demonstrated experimentally, see Ref.\,\cite{Korobko_PRL_131_143603_2023}.

In the current paper, we show that using the intracavity squeezing of the probe mode, it is possible to completely eliminate the SPM effect in the QND measurement scheme considered in Refs.\,\cite{21a1BaMaKhStIlSaLeBi, 22a1BaSaKh} and obtain the sensitivity approaching the single photon one. In addition, we provide here the rigorous quantitative analysis of this scheme, taking into account the finite bandwidths of the signal and probe modes, as well as the input, output and intracavity optical losses (in the works \cite{21a1BaMaKhStIlSaLeBi, 22a1BaSaKh}, the bandwidths of the modes were modelled by introducing the finite duration of interaction $\tau$ of these modes).

The paper is organized as follows. In Sec.\,\ref{sec:scheme}, we describe the QND measurement scheme considered here and introduce the main notations. In Sec.\ref{sec:in_out}, we derive the linearized input/output relations for the signal and probe beams. In Sec.\,\ref{sec:S}, we calculate the effective noise spectral densities that describe the measurement imprecision for the intracavity, input, and output values of the amplitude of the signal beam. In Sec.\,\ref{sec:estimates} we provide the sensitivity estimates for our scheme. In Sec.\,\ref{sec:conclusion} we briefly summarize the main results of the paper.

\section{Optical scheme and main notations}\label{sec:scheme}

\begin{figure*}
  \includegraphics[width=\textwidth]{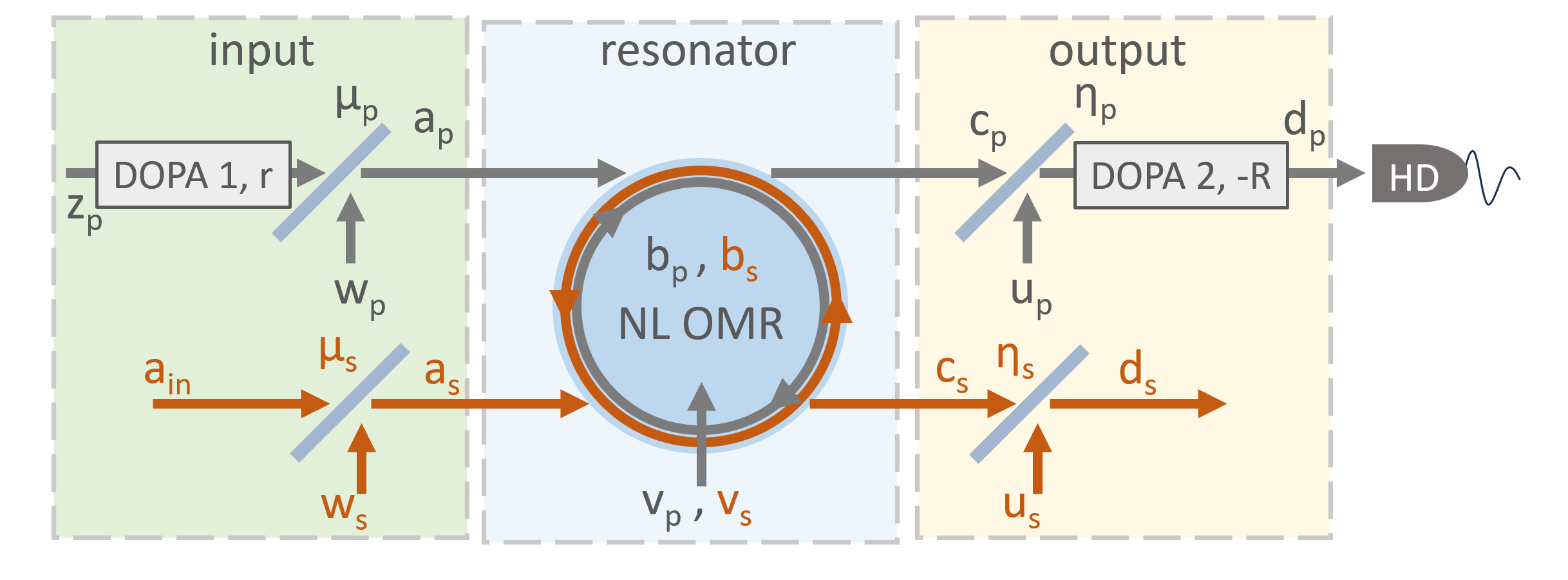} \\
  \caption{The scheme of the QND measurement. NL OMR - nonlinear optical microresonator, DOPA - degenerate optical parametric amplifier, HD - homodyne detector. The beamsplitters model the optical losses.}\label{fig:scheme}
\end{figure*}

The measurement scheme that we consider in this paper is shown in Fig.\,\ref{fig:scheme}. Here the signal and probe beams, denoted by the subscripts $s$ and $p$, respectively, are injected into the corresponding optical modes of the optical microresonator and interact there using the Kerr nonlinearity. We assume that the phase quadrature of the probe beam is prepared in a squeezed quantum state using the degenerate optical parametric amplifier DOPA 1. The phase quadrature of the corresponding output beam is anti-squeezed by the second degenerate optical parametric amplifier DOPA 2 and then measured by the homodyne detector HD.

In addition, we assume that the probe mode is squeezed inside the microresonator using additional parametric pump (not shown in Fig.\,\ref{fig:scheme}). The squeezing can be implemented using either three-wave mixing or four-wave mixing processes, with the same net result. In the former case, the quadratic ($\chi^{(2)}$) nonlinearity should be present in the microresonator, in addition to the Kerr nonlinearity. In the latter one, just the Kerr nonlinearity is sufficient. Here we, just to be specific, assume the first option.

We take into account the intracavity losses in both signal and probe modes, as well as the losses in the input and output paths of these modes (including photodetection inefficiency). We describe the input and output losses by means of imaginary beamsplitters model, see Ref.\,\cite{Leonhardt_PRA_48_4598_1993} and Fig.\,\ref{fig:scheme}. The power reflectivities of the input ``beamsplitters'' are denoted by $\mu_s$, $\mu_p$, and of the output ones --- by $\eta_s$, $\eta_p$.

Our analysis here is based on the formalism of the two-photon quadratures operators \cite{Caves1985, Schumaker1985} that are defined for any annihilation operator $\hat{a}$ as follows:
\begin{equation}\label{two_photon}
  \hat{a}^c(\Omega) = \frac{\hat{a}(\Omega) + \hat{a}^\dag(-\Omega)}{\sqrt{2}}\,, \quad
  \hat{a}^s(\Omega) = \frac{\hat{a}(\Omega) - \hat{a}^\dag(-\Omega)}{i\sqrt{2}}\,,
\end{equation}
Here $\Omega$ is the sideband frequency (the detuning from the respective carrier frequency) and the superscripts $c$, $s$ stand for the ``cosine'' and ``sine'' quadratures.

We denote the annihilation operators for the signal and probe fields as follows (see Fig.\,\ref{fig:scheme}): $\hat{a}_{s,p}$ --- the effective (that is after the input losses) input fields; $\hat{b}_{s,p}$ --- the intracavity fields; $\hat{c}_{s,p}$ --- the effective (before the output losses) output fields; $\hat{d}_{s,p}$ --- the output fields, taking into account the losses and the output anti-squeezing. We denote the annihilation operators of the effective vacuum noise fields, associated with the optical losses, as follows:  $\hat{w}_{s,p}$ --- the losses in the input paths; $\hat{v}_{s,p}$ --- the intracavity losses;  $\hat{u}_{s,p}$ --- the losses in the output paths.

For the particular case of the vacuum fields, the double-sided (that is defined for the frequencies $\Omega$ varying form $-\infty$ to $\infty$) spectral densities of the corresponding quadratures are equal to $1/2$.

\section{Input/output relations}\label{sec:in_out}

The scheme shown in Fig.\,\ref{fig:scheme} can be described by the following Hamiltonian:
\begin{multline}\label{hamilt}
  \frac{\hat H}{\hbar}
  = \sum_{x=s,p}\Bigl(
        \omega_x\hat b_x^{\dagger}\hat b_x
        - \frac{\gamma_S}{2}\hat b_x^{\dagger 2}\hat b_x^2
      \Bigr)
      - \gamma_X\hat b_p^{\dagger}\hat b_s^{\dagger}\hat b_p\hat b_s \\
      + \frac{ik}{2}(
            \hat b_p^{\dagger 2}e^{-2i\omega_p' - i\phi}
            - \hat b_p^2e^{2i\omega_p' + i\phi}
          )
     + \text{coupling terms} \,,
\end{multline}
where $\omega_{p,s}$ are the eigen frequencies of the signal and probe modes, $\gamma_{S,X}$ are the coefficients of the SPM ($S$) and XPM ($X$) interactions (see {\it e.g.} Eq.\,(61) in Ref.\,\cite{20a1BaKhStMaBi}), $k$ is the strength of the intracavity parametric excitation of the probe mode, $2\omega_p'$ is the frequency of the parametric pump, and $\phi$ is its phase. The ``coupling terms'' describe the intracavity losses and the coupling of the signal on probe modes with the input and output paths, see details in \eg Sec.\,III\,B of the review \cite{Aspelmeyer_RMP_86_1391_2014}.

The corresponding Heisenberg-Langevin equations of motion are derived in the Appendix \ref{app:io_eqs}. The outline of this derivation is the following. First, we switch to the rotating wave picture:
\begin{equation}\label{RWP}
  \hat{b}_{s,p}(t) \to \hat{b}_{s,p}(t)e^{-i\omega_{s,p}'t} \,,
\end{equation}
where
\begin{subequations}\label{dressed}
  \begin{gather}
    \omega_s' = \omega_s - \gamma_SN_s  - \gamma_XN_p \,, \\
    \omega_p' = \omega_p - \gamma_SN_p  - \gamma_XN_s
  \end{gather}
\end{subequations}
are the ``dressed'' eigen frequencies,
\begin{equation}
  N_{s,p} = |\beta_{s,p}|^2
\end{equation}
are the mean photon numbers in the respective modes, and $\beta_s$, $\beta_p$ are the classical amplitudes (the mean values) of $\hat{b}_s$, $\hat{b}_p$.

Then, we detach the classical amplitudes from the quantum fluctuations:
\begin{subequations}\label{linear}
  \begin{equation}
    \hat{b}_{s,p}(t) \to \beta_{s,p} + \hat{b}_{s,p}(t)
  \end{equation}
  (and in similar way for all other field operators) and keep only linear in the quantum fluctuations terms in the equations of motion. Without loss of generality, we assume that
  \begin{equation}\label{Re_betas}
    \Im\beta_s = \Im\beta_p = 0 \,.
  \end{equation}
\end{subequations}
In this case, the cosine quadratures of $\hat{b}_{s,p}$ (and of all other annihilation operators that appear below) describe the amplitude fluctuations of the respective fields, and the sine quadratures describe the phase fluctuations:
\begin{subequations}\label{b_n_phi}
  \begin{gather}
    \delta\hat{N}_{s,p} \approx \sqrt{2}\beta_s\hat{b}^c_{s,p} \,,\label{b_ampl} \\
    \hat{\phi}_{s,p}  \approx \frac{\hat{b}^s_{s,p}}{\sqrt{2}\beta_{s,p}} \,.\label{b_phi}
  \end{gather}
\end{subequations}

In the Fourier picture, the resulting linearized Heisenberg-Langevin equations have the following form:
\begin{subequations}\label{io_eqs}
  \begin{equation}
    \ell_s(\Omega)\hat b_s^c(\Omega) = \sqrt{2\kappa_s'}\hat a_{s}^c(\Omega)
      + \sqrt{2\kappa_s''}\hat v_s^c(\Omega) \,, \label{io_eqs_a}
  \end{equation}
  \begin{multline}
    \ell_s(\Omega)\hat b_s^s(\Omega) = B_{ss}\hat b_s^c(\Omega) + B_{sp}\hat b_p^c(\Omega)
      + \sqrt{2\kappa_s'}\hat a_{s}^s(\Omega) \\
      + \sqrt{2\kappa_s''}\hat v_s^s(\Omega) \,,
  \end{multline}
  \begin{equation}
    [\ell_p(\Omega) - k_c]\hat b_p^c(\Omega) + k_s\hat b_p^s
      = \sqrt{2\kappa_p'}\hat a_{p}^c(\Omega) + \sqrt{2\kappa_p''}\hat v_p^c(\Omega) \,,
  \end{equation}
  \begin{multline}
    [\ell_p(\Omega) + k_c]\hat b_p^s +(k_s - B_{pp})\hat b_p^c(\Omega)
      = B_{sp}\hat b_s^c(\Omega) \\
      + \sqrt{2\kappa_p'}\hat a_{p}^s(\Omega) + \sqrt{2\kappa_p''}\hat v_p^s(\Omega) \,.
      \label{io_eqs_d}
  \end{multline}
\end{subequations}
Here
\begin{equation}
  B_{ss} = 2\gamma_S\beta_s^2 \,, \quad B_{pp} = 2\gamma_S\beta_p^2 \,, \quad
  B_{sp} = 2\gamma_X\beta_s\beta_p
\end{equation}
are the rescaled SPM and XPM factors,
\begin{equation}\label{ell}
  \ell_{s,p}(\Omega) = -i\Omega + \kappa_{s,p} \,,
\end{equation}
\begin{equation}
  \kappa_{s,p} = \kappa_{s,p}' + \kappa_{s,p}''
\end{equation}
are the bandwidths of the signal and probe modes, $\kappa_{s,p}'$ are their parts originating from the coupling with the input fields, $\kappa_{s,p}''$ are the parts originating from the internal losses in the respective modes, and
\begin{equation}
  k_c = k\cos\phi \,, \quad k_s = k\sin\phi \,.
\end{equation}
The corresponding quadratures of the effective (lossless) output beams are equal to (see \eg review.\,\cite{Aspelmeyer_RMP_86_1391_2014}):
\begin{subequations}
  \begin{gather}
    \hat{c}_s^{c,s}(\Omega) = \sqrt{2\kappa_s'}\hat b_s^{c,s}(\Omega)
      - \hat a_s^{c,s}(\Omega) \,, \label{c_s} \\
    \hat{c}_p^{c,s}(\Omega) = \sqrt{2\kappa_p'}\hat b_p^{c,s}(\Omega)
      - \hat a_p^{c,s}(\Omega) \,. \label{c_p}
  \end{gather}
\end{subequations}

It follows from Eqs.\,\eqref{b_n_phi} that the term in the L.H.S. of Eq.\,\eqref{io_eqs_d} proportional to $B_{pp}$ is responsible for the undesirable SPM effect. Evidently, this term can be canceled by setting
\begin{equation}
  k_s = B_{pp} \,,
\end{equation}
giving the following equation for the sine quadrature of the probe mode:
\begin{equation}\label{b_p^s}
  \hat b_p^s(\Omega) = \frac{1}{\ell_p(\Omega) + k_c}\Bigl[
      B_{sp}\hat b_s^c(\Omega) + \sqrt{2\kappa_p'}\hat a_{p}^s(\Omega)
      + \sqrt{2\kappa_p''}\hat v_p^s(\Omega)
    \Bigr] .
\end{equation}
This equation will be used extensively in the next section.

\section{Noise spectral densities}\label{sec:S}

\subsection{Intracavity field}\label{sec:intra}

In this subsection, we calculate the spectral density that describes the measurement imprecision for the amplitude (cosine) quadrature of the intracavity field.

Taking into account the anti-squeezing and the output losses (including the detection inefficiency), the sine quadrature of the output beam can be presented as follows:
\begin{equation}\label{d_p^s}
  \hat d_p^s(\Omega) = \sqrt{\eta_p}\hat c_p^s(\Omega)e^R + \sqrt{1-\eta_p}\hat u_p^s(\Omega)
    \,,
\end{equation}
where $e^R$ is the anti-squeeze factor.
Combining Eqs.\,\eqref{c_p}, \eqref{b_p^s}, and \eqref{d_p^s}, we obtain that:
\begin{equation}\label{probe_out}
  \hat d^s_{p}(\Omega)
  = G_{\rm intr}(\Omega)\bigl[\hat b_s^c(\Omega) + \hat b_{\rm intr}^c(\Omega)\bigr]
\end{equation}
where
\begin{equation}
  G_{\rm intr}(\Omega) = \frac{\sqrt{2\eta_p\kappa_p'}B_{sp}e^R}{\ell_p(\Omega) + k_c}
\end{equation}
is the common gain factor for the intracavity field measurement,
\begin{multline}\label{b_intr_c}
  \hat b_{\rm intr}^c(\Omega) = \frac{1}{\sqrt{2\kappa_p'}B_{sp}}\bigl\{
      (\kappa_p' - \kappa_p'' - k_c + i\Omega)\hat a_{p}^s(\Omega) \\
      + 2\sqrt{\kappa_p'\kappa_p''}\hat v_p^s(\Omega)
      + \epsilon[\ell_c(\Omega) + k_c]\hat u_p^s(\Omega)e^{-R}
    \bigr\}
\end{multline}
is the normalized intracavity quantum noise, and
\begin{equation}
  \epsilon = \sqrt{\frac{1-\eta_p}{\eta_p}} \,.
\end{equation}
is the normalized output loss factor of the probe mode.

We assume that the input noise $\hat{a}_p$ is prepared in the squeezed state by means of the input parametric amplifier {\sf DOPA 1}, and take into account the input losses:
\begin{equation}
  \hat a_p^s =\sqrt{\mu_p}e^{-r_0}\hat z_p^s+ \sqrt{1-\mu_p}\hat w_p^s,
\end{equation}
where $\hat{z}_p$ is the vacuum field and $r_0$ is the ``raw'' squeeze factor. In this case, spectral density of $\hat b_{\rm intr}^c$ is equal to
\begin{multline}\label{Sintr_raw}
  S_{\rm intr}(\Omega) = \frac{1}{4\kappa_p'B_{sp}^2}\bigl[
      (e^{-2r} + \epsilon^2e^{-2R})\Omega^2
      + (\kappa_p' - \kappa_p'' - k_c)^2e^{-2r} \\
      + 4\kappa_p'\kappa_p'' + \epsilon^2(\kappa_p + k_c)^2e^{-2R}
    \bigr] ,
\end{multline}
where
\begin{equation}
  e^{-2r} = 1 - \mu_p + \mu_pe^{-2r_0}
\end{equation}
and $r$ is the squeeze factor degraded by the losses.

The minimum of the spectral density \eqref{Sintr_raw} is provided by
\begin{equation}\label{k_c_opt}
  k_c = \frac{(\kappa_p' - \kappa_p'')e^{-2r} - \epsilon^2\kappa_pe^{-2R}}
    {e^{-2r} + \epsilon^2e^{-2R}} \,.
\end{equation}
and is equal to
\begin{equation}\label{S_intr}
  S_{\rm intr}(\Omega) = \frac{1}{4\gamma_X^2N_pN_s}\biggl(
      \frac{\epsilon^2e^{-2R} + e^{-2r}}{4\kappa_p'}\Omega^2
      + \frac{\epsilon^2\kappa_p'}{e^{2R} + \epsilon^2e^{2r}} + \kappa_p''
    \biggr) .
\end{equation}

\subsection{Input field}\label{sec:meas}

Then, let us calculate the spectral density that describes the imprecision of the input field measurement.

Taking into account the input losses, the effective input field of the signal mode can be presented as follows:
\begin{equation}\label{a_in}
  \hat{a}_s(\Omega) = \sqrt{\mu_s}\hat{a}_{\rm in}(\Omega) + \sqrt{1-\mu_s}\hat{w}_s(\Omega)
    \,.
\end{equation}
where $\hat{a}_{\rm in}$ is the real incident field, see Fig.\,\ref{fig:scheme}.

Combining this equation with Eqs.\,\eqref{io_eqs_a} and \eqref{probe_out}, we obtain that
\begin{equation}\label{d_p_meas}
  \hat{d}_p^s(\Omega)
  = G_{\rm in}(\Omega)\bigl[\hat{a}^c_{\rm in}(\Omega) + \hat{a}_{\rm in}^c(\Omega)] ,
\end{equation}
where
\begin{equation}\label{G_meas}
  G_{\rm in}(\Omega) = \frac{\sqrt{2\mu_s\kappa_s'}}{\ell_s(\Omega)}G_{\rm intrr}(\Omega)
\end{equation}
is the common gain factor for the input field measurement and
\begin{equation}\label{a_meas_c}
   \hat{a}_{\rm in}^c(\Omega)
   = \frac{\ell_s(\Omega)}{\sqrt{2\mu_s\kappa_s'}}\hat{b}_{\rm intr}^c(\Omega)
    + \sqrt{\frac{1-\mu_s}{\mu_s}}\hat{w}_s^c(\Omega)
    + \sqrt{\frac{\kappa_s''}{\mu_s\kappa_s'}}\hat{v}_s^c(\Omega)
\end{equation}
is the effective measurement noise with the spectral density equal to
\begin{equation}\label{S_meas}
  S_{\rm in}(\Omega) = \frac{1}{2\mu_s}\biggl[
      \frac{\Omega^2 + \kappa_s^2}{\kappa_s'}S_{\rm intr}(\Omega)
      + \frac{\kappa_s''}{\kappa_s'} + 1 - \mu_s
    \biggr] .
\end{equation}

\subsection{Output field}\label{sec:prep}

Finally, let us calculate the spectral density that describes the imprecision of the output field measurement, that is, the quality of preparation of the output quantum state.

Taking into account of the optical losses, the output field of the signal mode can be presented as follows:
\begin{equation}
  \hat{d}_s^c(\Omega)
  = \sqrt{\eta_s}\hat{c}_s^c(\Omega) + \sqrt{1-\eta_s}\hat{u}_s^c(\Omega) \,.
\end{equation}
Combining this equation with Eqs.\,\eqref{io_eqs_a} and \eqref{c_s}, we obtain:
\begin{multline}\label{d_s_c}
  \hat{d}_s^c(\Omega)
  = \sqrt{\frac{\eta_s}{2\kappa_s'}}\bigl[
        (i\Omega + \kappa_s' - \kappa_s'')\hat{b}_s^c(\Omega)
        + \sqrt{2\kappa_s''}\hat{v}_s^c(\Omega)
      \bigr] \\
    + \sqrt{1-\eta_s}\hat{u}_s^c(\Omega) \,.
\end{multline}
Solving this equation for $\hat{b}_s^c$:
\begin{equation}
  \hat{b}_s^c = \frac{1}{i\Omega + \kappa_s' - \kappa_s''}\Biggl(
      \sqrt{\frac{2\kappa_s'}{\eta_s}}\hat{d}_s^c - \sqrt{2\kappa_s''}\hat{v}_s^c
      - \sqrt{2\kappa_s'\frac{1-\eta_s}{\eta_s}}\hat{u}_s^c
    \Biggr) .
\end{equation}
and substituting this value into Eq.\,\eqref{probe_out}, we obtain that
\begin{equation}\label{d_p_prep}
  \hat{d}_p^s
  = G_{\rm out}(\Omega)\bigl[\hat{d}_s^c(\Omega) + \hat{d}_{\rm out}^c(\Omega)\bigr] \,,
\end{equation}
where
\begin{equation}\label{G_prep}
  G_{\rm prep}(\Omega)
  = \frac{\sqrt{2\kappa_s'/\eta_s}}{i\Omega + \kappa_s' - \kappa_s''}G_{\rm intr}(\Omega)
\end{equation}
is the common gain factor for the output field measurement and
\begin{equation}\label{d_prep_c}
   \hat{d}_{\rm out}^c
   = \frac{i\Omega + \kappa_s' - \kappa_s''}{\sqrt{2\kappa_s'/\eta_s}}\hat{b}_{\rm intr}^c
     - \sqrt{\eta_s\frac{\kappa_s''}{\kappa_s'}}\,\hat{v}_s^c
       - \sqrt{1-\eta_s}\hat{u}_s^c \,.
\end{equation}
is the corresponding noise with the spectral density equal to
\begin{equation}\label{S_prep}
  S_{\rm out}(\Omega) = \frac{\eta_s}{2}\biggl[
      \frac{\Omega^2 + (\kappa_s' - \kappa_s'')^2}{\kappa_s'}S_{\rm intr}(\Omega)
      + \frac{\kappa_s''}{\kappa_s'} + \frac{1-\eta_s}{\eta_s}
    \biggr] .
\end{equation}

\section{Estimates of the sensitivity}\label{sec:estimates}

The spectral densities calculated in the previous section, see Eqs.\,\eqref{S_intr}, \eqref{S_meas}, \eqref{S_prep}, depend on the running frequency $\Omega$. Therefore, the rigorous calculation of the corresponding photon number measurement errors requires the knowledge of the exact form of signal pulse shape. This task exceeds the scope of this article. At the same time, approximate sensitivity estimates can be obtained by replacing $\Omega$ with some characteristic value. In the practically important case of a smooth (\eg Gaussian) shape of the signal, the value $1/\tau$, where $\tau$ is the signal pulse duration, can be used as this frequency, as it was done de facto in the work \cite{22a1BaSaKh}. Also following that work, we assume that the signal pulse duration matches the bandwidths of both the signal and probe modes:
\begin{equation}
  \frac{1}{\tau} \approx \kappa_p \approx \kappa_s \,.
\end{equation}
Finally, we assume that the optical losses in our scheme are small:
\begin{equation}
  \kappa_{s,p}' \ll \kappa_{p,s} \,, \quad 1-\mu_{s,p} \ll 1 \,, \quad 1-\eta_{s,p} \ll 1
    \,.
\end{equation}
In this case, Eqs.\,\eqref{S_intr}, \eqref{S_meas}, \eqref{S_prep} can be simplified as follows:
\begin{gather}
  S_{\rm intr} = \frac{\tau}{4\Gamma_X^2N_pN_s}\biggl(
      \frac{\epsilon^2e^{-2R} + e^{-2r}}{4}
      + \frac{\epsilon^2}{e^{2R} + \epsilon^2e^{2r}} + \kappa_p''\tau
    \biggr) \label{S0_intr} ,  \\
  S_{\rm in} = \frac{S_{\rm intr}}{\tau} + \frac{\epsilon_{\rm in}^2}{2}\,,
    \label{S0_meas}\\
  S_{\rm out} =
    \frac{S_{\rm intr}}{\tau} + \frac{\epsilon_{\rm out}^2}{2} \,, \label{S0_prep}
\end{gather}
where
\begin{equation}
  \Gamma_X = \tau\gamma_X \,.
\end{equation}
and
\begin{subequations}
  \begin{gather}
    \epsilon_{\rm in}^2 = \kappa_s''\tau + 1 - \mu_s \,, \\
    \epsilon_{\rm out}^2 = \kappa_s''\tau + 1 - \eta_s
  \end{gather}
\end{subequations}
are the ``in'' and ``out'' loss factors for the signal beam.

It is instructive to compare these results with the ones obtained in Ref.\,\cite{22a1BaSaKh}. In order to facilitate this comparison, we recast Eqs.\,\eqref{S0_intr}-\eqref{S0_prep} in terms of photon number measurement errors:
\begin{multline}
  (\Delta N)_{\rm intr}^2 = \frac{2N_sS_{\rm intr}}{\tau} \\
  = \frac{1}{2\Gamma_X^2N_p}\biggl(
      \frac{\epsilon^2e^{-2R} + e^{-2r}}{4}
      + \frac{\epsilon^2}{e^{2R} + \epsilon^2e^{2r}} + \kappa_p''\tau
    \biggr) \label{D2N_intr} ,
\end{multline}
\begin{gather}
  (\Delta N)_{\rm in}^2 = 2N_sS_{\rm in}
    = (\Delta N)_{\rm intr}^2 + \epsilon_{\rm in}^2N_s \,, \label{D2N_meas} \\
  (\Delta N)_{\rm out}^2 = 2N_sS_{\rm out}
    = (\Delta N)_{\rm intr}^2 + \epsilon_{\rm out}^2N_s \label{D2N_prep} \,.
\end{gather}

Consider first the intracavity measurement error \eqref{D2N_intr}. The first term in this equation, up to the numeric factor $1/2$ (which can be explained by the more rigorous approach used in the current paper), coincides with the corresponding one in Eq.\,(36) of \cite{22a1BaSaKh}. The last term is absent in Eq.\,(36) of \cite{22a1BaSaKh} because the internal losses were neglected in that work. At the same time, the second term differs radically. In Eq.\,\eqref{D2N_intr} it scales with the photon number in the probe mode as $1/N_p$, but in Eq.\,(36) of \cite{22a1BaSaKh} --- as $N_p$. Therefore, while in the latter case the sensitivity is limited by a value that is independent on $N_p$ (see Eq.\,(40) of \cite{22a1BaSaKh}), in the former case, arbitrarily high sensitivity can be acheved by increasing $N_p$.

It is easy to see that the following balanced values of the squeeze factors:
\begin{equation}
  e^{R-r} = \epsilon \,
\end{equation}
provide the minimum of \eqref{D2N_intr}, equal to
\begin{equation}\label{optD2N}
  (\Delta N)_{\rm intr}^2 = \frac{\epsilon e^{-r-R} + \kappa_p''\tau}{2\Gamma_X^2N_p} \,.
\end{equation}

For the numerical estimates, we use the same main parameters as in Ref.\,\cite{22a1BaSaKh}. We assume that a microresonator made of $\text{CaF}_2$ with an intrinsic $Q$-factor $Q=3\times10^{11}$ \cite{07a1SaIlMa} is used. We also assume that $\tau$ corresponds to the loaded $Q$-factor $Q\approx10^{10}$, giving $\Gamma_X\approx10^{-5}$. Taking into account the recent advances in the preparation of squeezed quantum states (see, \eg, Refs.\,\cite{Vahlbruch_PRL_117_110801_2016, Meylahn_PRL_129_121103_2022}), it is easy to see that the numerator of Eq.\,\eqref{optD2N} can be made smaller than 0.1, giving that
\begin{equation}\label{estD2N}
  (\Delta N)_{\rm intr}^2 \sim \frac{10^9}{N_p} \,.
\end{equation}
This means that using about $10^9$ photons in the probe mode, single photon sensitivity can be achieved for the number of photons in the signal mode

Now consider the meansurement errors of the incident and outgoing photon numbers. It follows from Eqs.\,\eqref{D2N_meas} and \eqref{D2N_prep}, that in order to achieve the single photon sensitivity, the mean photon number in the probe mode has to be sufficiently small:
\begin{equation}
  N_s < \frac{1}{\epsilon^2_{\rm in,\,out}}  \,.
\end{equation}
Assume the moderately optimisitic value
\begin{equation}\label{eps2s}
  \epsilon^2_{\rm in,\,out} \sim 0.1 \,.
\end{equation}
In this case, the single photon sensitivity can be achieved only if
\begin{equation}
  N_s \lesssim 10 \,.
\end{equation}

At the same time, another important threshold for the photon number measurement error is known that applies a more relaxed limitation on the photon number. It was shown in Refs.\,\cite{Bondurant_PRD_30_2548_1984, Kitagawa_PRA_34_3974_1986}, that all pure quantum states with the photon-number uncertainty satisfying the following inequality (we omit for brevity the numerical factors of order of unity):
\begin{equation}
  \Delta N < N^{1/3} \,,
\end{equation}
are non-Gaussian states, described by negative-valued Wigner quasiprobability distributions. These quantum states could be very interesting in particular, for the continuous-variable quantum computing.

It is easy to see that this constraint translates to the following limitation on the mean photon number:
\begin{equation}
  N_s \lesssim \frac{1}{\epsilon^6_{\rm in,\,out}} \,.
\end{equation}
In the case of \eqref{eps2s}, this means that the QND measurement scheme considered here allows to verify and prepare bright non-Gaussian quantum states with the mean photon number up to $\sim10^3$.

\section{Conclusion}\label{sec:conclusion}

In this work, we extended the analysis of considered in Refs.\,\cite{22a1BaSaKh} scheme of quantum non-demolition measurement of optical quanta that employs a resonantly enhanced Kerr nonlinearity in optical microresonators. It has been shown in that work that the main limiting factor for the sensitivity of this scheme is the interplay of optical losses and the non-linear self phase modulation (SPM) effect.

Here we showed that using the intracavity squeezing of the probe beam, it is possible to cancel the SPM effect. Using linearized Heisenberg-Langevin equations, we calculated the corresponding sensitivity limits imposed by the input, output and intracavity optical losses in both probe and signal modes.

We showed that the sensitivity of the considered scheme is limited only by the available power in the pump beam and losses in the signal beam. Our estimates show that, using the best optical microresonators currently available, the single-photon sensitivity for the intracavity photon number can be achieved, and generation and verification of bright non-Gaussian quantum states with a mean photon number up to $\sim10^3$ is feasible. Therefore, the QND measurement scheme considered here could be interesting for the quantum information processing tasks, in partcular, for the continuous-variable quantum computing.

\section{Acknowledgment}

This work was supported by the Russian Science Foundation (project 20-12-00344). The research of Sec.\,\ref{sec:estimates} was supported by Rosatom in the framework of the Roadmap for Quantum computing (Contract No. 868-1.3-15/15-2021 dated October 5, 2021 and Contract No.P2154 dated November 24, 2021).

\appendix

\section{Derivation of Eqs.\,\eqref{io_eqs}}\label{app:io_eqs}

In the Heisenberg picture, the following equations of motion can be derived from the Hamiltonian \eqref{hamilt}:
\begin{subequations}\label{eqs_raw}
  \begin{equation}
    \fulld{\hat{b}_s}{t} = -i\omega_s\hat{b}_s
      + i\gamma_S\hat{b}_s^\dag\hat{b}_s^2 + i\gamma_X\hat{b}_p^\dag\hat{b}_p\hat{b}_s
      + \text{coupling terms} ,
  \end{equation}
  \begin{multline}
    \fulld{\hat{b}_p}{t} = -i\omega_p\hat{b}_p
      + i\gamma_S\hat{b}_p^\dag\hat{b}_p^2 + i\gamma_X\hat{b}_p\hat{b}_s^\dag\hat{b}_s
      + k\hat{b}_p^\dag e^{-2i\omega_p't-i\phi} \\ + \text{coupling terms} .
  \end{multline}
\end{subequations}
Switch to the rotating wave picture, see Eqs.\,\eqref{RWP}:
\begin{subequations}\label{eqs_RWP}
  \begin{multline}
    \fulld{\hat{b}_s}{t} = i(\omega_s' - \omega_s)\hat{b}_s
      + i\gamma_S\hat{b}_s^\dag\hat{b}_s^2 + i\gamma_X\hat{b}_p^\dag\hat{b}_p\hat{b}_s \\
      + \text{coupling terms} \,,
  \end{multline}
  \begin{multline}
    \fulld{\hat{b}_p}{t} = i(\omega_p' - \omega_p)\hat{b}_p
      + i\gamma_S\hat{b}_p^\dag\hat{b}_p^2 + i\gamma_X\hat{b}_p\hat{b}_s^\dag\hat{b}_s
      + k\hat{b}_p^\dag e^{-i\phi} \\ + \text{coupling terms} .
  \end{multline}
\end{subequations}
Using the substitution \eqref{linear}, keeping only linear in $\hat{b}_{c,s}$, $\hat{b}_{c,s}^\dag$ terms, and assuming Eqs.\, \eqref{dressed}, we obtain:
\begin{subequations}
  \begin{equation}
    \fulld{\hat{b}_s}{t}
      = \frac{i}{2}[B_{ss}(\hat{b}_s + \hat{b}_s^\dag) + B_{sp}(\hat{b}_p + \hat{b}_p^\dag)]
         + \text{coupling terms} \,,
  \end{equation}
  \begin{multline}
    \fulld{\hat{b}_p}{t}
      = \frac{i}{2}
          [B_{pp}(\hat{b}_p + \hat{b}_p^\dag) + B_{sp}(\hat{b}_s + \hat{b}_s^\dag)]
        + k\hat{b}_p^\dag e^{-i\phi} \\ + \text{coupling terms} ,
  \end{multline}
\end{subequations}
or, in the Fourier picture:
\begin{subequations}\label{eqs_lin_Omega}
  \begin{equation}
    -i\Omega\hat{b}_s
      = \frac{i}{2}[B_{ss}(\hat{b}_s + \hat{b}_s^\dag) + B_{sp}(\hat{b}_p + \hat{b}_p^\dag)]
         + \text{coupling terms} \,,
  \end{equation}
  \begin{multline}
    -i\Omega\hat{b}_p
      = \frac{i}{2}
          [B_{pp}(\hat{b}_p + \hat{b}_p^\dag) + B_{sp}(\hat{b}_s + \hat{b}_s^\dag)]
        + k\hat{b}_p^\dag e^{-i\phi} \\ + \text{coupling terms} ,
  \end{multline}
\end{subequations}
where
\begin{equation}
  B_{ss} = 2\gamma_S\beta_s^2 \,, \quad B_{pp} = 2\gamma_S\beta_p^2 \,, \quad
  B_{sp} = 2\gamma_X\beta_s\beta_p \,.
\end{equation}
In the two-photon quadratures notations \eqref{two_photon}, Eqs.\,\eqref{eqs_lin_Omega} have the following form:
\begin{subequations}\label{eqs_lin_raw}
  \begin{gather}
    -i\Omega\hat{b}_s^c = \text{coupling terms} , \\
    -i\Omega\hat{b}_s^s = B_{ss}\hat{b}_s^c + B_{sp}\hat{b}_p^c + \text{coupling terms},
      \\
    -i\Omega\hat{b}_p^c = k(\hat{b}_p^c\cos\phi - \hat{b}_p^s\sin\phi)
      + \text{coupling terms} ,
  \end{gather}
  \begin{multline}
    -i\Omega\hat{b}_p^s = B_{pp}\hat{b}_p^c + B_{sp}\hat{b}_s^c
      - k(\hat{b}_p^c\sin\phi + \hat{b}_p^s\cos\phi) \\ + \text{coupling terms} .
  \end{multline}
\end{subequations}
Finally, adding to these equations the explicit expressions for the ``coupling terms'', see \eg Sec.\,III\,B of \cite{Aspelmeyer_RMP_86_1391_2014}, we obtain Eqs.\,\eqref{io_eqs}.


\providecommand{\noopsort}[1]{}\providecommand{\singleletter}[1]{#1}%

\end{document}